\documentstyle[11pt,aaspp4,flushrt]{article}

\received{\underline{Oct 09, 1997}}
\accepted{\underline{Nov 18, 1997}}

\slugcomment{\hfill To appear in ApJ Letters}

\lefthead{Tsvetanov et al.}
\righthead{Optical variability of M87}

\def\HaNii{{H$\alpha$+[\ion{N}{2}]}\/}

\def\ergs.cm2.s{{ergs~cm$^{-2}$~s$^{-1}$}\/}
\def\as{{$''\!\!$.}\/}

\def\deg{{$^{\circ}$}\/}

\begin{document}

\title{M87: A MISALIGNED BL LAC?}

\author{\sc Zlatan I.\ Tsvetanov\altaffilmark{1}, 
	    George F.\ Hartig\altaffilmark{2},  
	    Holland C.\ Ford\altaffilmark{1,2}, 
	    Michael A.\ Dopita\altaffilmark{3},
	    Gerard A.\ Kriss\altaffilmark{1},  
	    Yichuan C.\ Pei\altaffilmark{1}, 
	    Linda L.\ Dressel\altaffilmark{4},  
            Richard J.\ Harms\altaffilmark{4}}

\altaffiltext{1}{Department of Physics and Astronomy, Johns Hopkins
	University, Baltimore, Maryland 21218, USA; zlatan@pha.jhu.edu} 
\altaffiltext{2}{Space Telescope Science Institute, 3700 San Martin Dr.,  
	Baltimore, MD 21218, USA}
\altaffiltext{3}{Mount Stromlo and Siding Spring Observatories, 
        Australian National University, Private Bag Weston Creek P.O.,
        ACT 2611, Australia; mad@mso.mso.anu.edu.au}
\altaffiltext{4}{RJH Scientific, Inc., 5904 Richmond Highway, Suite 401,  
	Alexandria, VA 22303, USA}

\begin{abstract}

The nuclear region of M87 was observed with the Faint Object
Spectrograph (FOS) on the {\it Hubble Space Telescope} ($HST$) at 6
epochs, spanning 18 months, after the $HST$ image quality was improved
with the deployment of the corrective optics (COSTAR) in December
1993.  From the FOS target acquisition data, we have established that
the flux from the optical nucleus of M87 varies by a factor $\sim$2 on
time scales of $\sim$2.5 months and by as much as 25\% over 3 weeks,
and remains unchanged ($\lesssim$ 2.5\%) on time scales of $\sim$ 1
day.  The changes occur in an unresolved central region $\lesssim$ 5
pc in diameter, with the physical size of the emitting region limited
by the observed time scales to a few hundred gravitational radii. The
featureless continuum spectrum becomes bluer as it brightens while
emission lines remain unchanged.  This variability combined with the
observations of the continuum spectral shape, strong relativistic
boosting and the detection of significant superluminal motions in the
jet, strongly suggest that M87 belongs to the class of BL Lac objects
but is viewed at an angle too large to reveal the classical BL Lac
properties.

\end{abstract}


\section{INTRODUCTION}

Variability is a fundamental property of Active Galactic Nuclei
(AGN). In fact, it was the rapid changes of the observed optical
continuum flux of quasars that led early researchers to the conclusion
that vast luminosities must be generated in a very small volume,
suggesting that accretion onto a massive black hole is the source of
the energy generation (Salpeter 1964, Zeldovich \& Novikov 1964).

It is now understood that in the BL Lac objects rapid variability is
the result of strong relativistic boosting in a jet oriented close to
the line of sight to the observer (for a review see Urry \& Padovani
1995 and references therein). The connection between BL Lacs and the
Fanaroff-Riley class I (FR~I) radio sources is now firmly established
(Padovani 1992). The flux in the extended radio structure of BL Lacs
lies in the same range as observed for the FR~I sources.

Apart from the variability, the fundamental properties of BL Lacs are:
(1) bright featureless continuum of synchrotron origin which peaks
typically at IR wavelengths for the radio-selected objects (the low
energy BL Lacs, or LBL) and in the UV/X-ray region for the X-ray
selected ones (high energy BL Lacs, HBL -- see Urry \& Padovani 1995
and references therein for a description of the LBL/HBL dichotomy);
(2) superluminal motions in the jets observed at radio frequencies
(Vermeulen \& Cohen 1994; Gabuzda et al.\ 1994); (3) BL Lacs are found
exclusively in elliptical galaxy hosts. Because FR~I radio galaxies
often exhibit with LINER-like nuclear spectra (e.g., Baum, Zirbel \&
O'Dea 1995), and BL Lacs are thought to be drawn from the same
population as FR~I (see Urry \& Padovani 1995), it is then likely that
some BL Lacs would also display LINER type activity if the beamed 
continuum radiation did not overwhelm the emission lines.

The galaxy M87 (NGC~4486) shares many of these properties. Although
M87 is 0.3 mag fainter than NGC 4472 (M49), it is the dominant galaxy
in the largest optical component of Virgo, Cloud A (Binggeli et al.\
1987), and is at the center of the corresponding extended hard X-ray
component that accounts for $\sim 70\%$ of Virgo's X-ray Luminosity
(Bohringer et al. 1994). M87 is presently the best example of a small
disk of ionized gas ($r \sim 100$ pc; Ford et al.\ 1994) fueling a
massive black hole $M_{\rm BH} \sim 2-3 \times 10^9~M_{\odot}$ (Harms
et al.\ 1994, Ford et al.\ 1996; Macchetto et al.\ 1997). The gaseous
disk has a classical LINER emission line spectrum, and its properties
in the UV establish it as a shock heated accretion disk (Dopita et
al.\ 1997).  The ``engine'' in the center produces the best known
example of an extragalactic optical synchrotron jet oriented at about
$\sim 30-35^{\circ}$ to our line of sight (e.g., Biretta 1994;
Bicknell \& Begelman 1996). In the X-rays significant variability has
recently been reported by Harris, Biretta \& Junor (1997).  The
broad-band spectrum of the nuclear source from the radio to the X-rays
is similar to that of the individual knots in the jet. The
radio-to-optical index, $\alpha_{\rm ro} = 0.81$, the
optical-to-X--ray index, $\alpha_{\rm ox} = 1.12$, and the optical--UV
spectral index is 1.46 (Biretta, Stern, \& Harris 1991; Boksenberg et
al.\ 1992), making it qualitatively similar to optically selected
blazars such as 3C~345 (Kollgaard 1994; Maraschi et al.\ 1986; Bregman
et al.\ 1986). In fact, these values of $\alpha_{\rm ro}$ and
$\alpha_{\rm ox}$ place M87 at the edge of the region of the
$\alpha_{\rm ro} - \alpha_{\rm ox}$ plane populated by radio-selected
BL Lacs and radio quasars (Padovani \& Giommi 1995). The jet displays
powerful relativistic boosting, strong internal shocks, and
significant superluminal motions (Biretta, Zhou \& Owen 1995). In
addition this jet is powering complex radio lobes with luminosity and
morphology typical of FR~I sources.

The proximity of M87, combined with independent measurements of the
central black hole mass and jet orientation, make it a primary target
for studying some of the fundamental properties of active galactic
nuclei. 

In this paper we present analysis of the FOS target acquisition (TA)
data for several observations of M87's central region. The data reveal
significant variations of the nonstellar nuclear source flux in the
optical wavelength region.  Throughout the paper we assume a distance
to M87 of 15 Mpc (Jacoby, Ciardullo \& Ford 1990) which simplifies
comparisons with results of the latest measurements of the central
black hole mass (Harms et al.\ 1994; Ford et al.\ 1996; Macchetto et
al.\ 1997), the energetics of the LINER disk (Dopita et al.\ 1997),
and the X-ray variability (Harris, Biretta \& Junor 1997).

\section{OBSERVATIONS}

As part of several GTO and GO programs, the nuclear region of M87 was
observed with the FOS on $HST$ at 6 epochs over 18 months, following
the introduction of COSTAR in December 1993. The spectra obtained at
each epoch were preceded by a careful target acquisition consisting of
a binary search (BS) followed by one or more peakup rasters, all using
the FOS red channel in camera mode. The photocathode of the FOS red
detector is sensitive to wavelengths in the 1600 -- 8000 \AA\ region,
with peak quantum efficiency of $\sim 25$\% between 2000 -- 4000
\AA. The binary search data comprise a series of one-dimensional
images obtained by electronically stepping the target field in the
3\farcs7 square aperture across the 1\farcs29 tall linear diode
array. For the peakup sequences, the 0\farcs26 diameter circular
aperture was stepped in a raster pattern over the target field
utilizing small angle maneuvers of the telescope. The two types of
target acquisition data yield corroborating evidence for strong
nuclear intensity variations.

Fig.~1 shows the BS images of the centered nucleus at four of the
epochs (the other two are not shown for clarity). Note that the
underlying galactic component remains constant to within the
measurement errors, while the nuclear source varies in intensity.  The
width of the nuclear profile remains constant at about 0.9 diode
widths, which is characteristic of an unresolved point source. The
insert in Fig.~1 shows the shape of the radial profile of the M87
nucleus inferred from one of the peakup series. Overplotted is a
scaled profile of a star, again illustrating that the nucleus of M87
is unresolved by the $HST$.  The characteristic size of the $HST$ PSF
at the FOS after the introduction of COSTAR is $\sim$60 mas FWHM
(Hartig, Crocker \& Ford 1994).  At a distance to M87 of 15 Mpc, the
upper limit on the size of the region producing the variability is
$\lesssim 5$ pc in diameter.

At each epoch we computed the nuclear source count rate when the
aperture was best centered on the nucleus. For the BS data we
integrated the flux above the background galactic contribution, which
was approximated by a constant and left free in the fitting. The
integration was restricted to a 1\farcs01 region centered on the peak
(see Fig.~1), which effectively represents the flux of the nucleus
measured in a 1\as01 $\times$ 1\farcs29 aperture. For the peakup
series we used the count rate observed at the best centered
position. The contribution from the host galaxy in the 0\farcs26
circular aperture used in the peakup series was estimated to be less
than a few \% and was neglected.  Both measurements are illustrated in
Fig.~2 as a function of time, with estimates of the error and the
level of the galactic component.  The flux ratio between the 0\farcs26
and 1\farcs01 $\times$ 1\farcs29 apertures is close to what is
expected for the relative transmission of a point source (Keyes et
al.\ 1995), and remains constant for all measurements. Fig.~2 shows
that the nuclear flux changes by a factor of 2 over approximately 2.5
months (78 days) and by as much as 25\% in 3 weeks. The flux of the
nucleus did not change to well within the 1$\sigma$ error bars (2.5\%)
for the two measurements obtained only 23 hours apart on 24 and 25 May
1995.
 
For two of the observational epochs a nuclear spectrum was obtained
through the 0\farcs26 circular aperture (Fig.~3).  In both cases the
spectrum is dominated by bright broad emission lines and a strong
non-thermal blue continuum, similar to other types of AGN such as
quasars.  The continuum is well described by a power law with spectral
index $\alpha = 1.4$ ($F(\nu) \sim \nu^{-\alpha}$), in agreement with
that of the unresolved nuclear source determined from $HST$ imaging
(Boksenberg et al.\ 1992).  The \ion{Na}{1} line seen in absorption
has the same redshift as M87 and is a neutral component of M87's
extended ISM.  No other absorption features are seen, indicating that
any contributions due to stellar light is small. The ratio of the two
spectra plotted in the lower panel of Fig.~3 shows two important
characteristics. First, the spectrum becomes bluer as it brightens.
This is in good agreement with properties of the radio-selected BL
Lacs which seem to display the same behavior in the IR/optical/UV band
(see Ulrich, Maraschi \& Urry 1997). The fractional change at
$\lambda$4600 \AA\ is approximately 50\% versus only 35\% at
$\lambda$6800 \AA. Second, the ratio at the emission line positions is
significantly lower, indicating that the observed changes are due to a
rise in the continuum level. This is best seen at the position of the
bright \HaNii\ complex which is less subject to a continuum
dilution. The ratio in that wavelength region drops to almost 1,
suggesting that emission line fluxes may indeed remain
unchanged. Higher signal-to-noise spectra, however, are needed to
investigate this in more detail.

\section{DISCUSSION}

The variability characteristic time scale is of particular interest
when studying the underlying phenomena. In practice, the time scale
depends on the definition, wavelength region, etc. To avoid ambiguity,
we adopt a working definition that the typical time scale for
variability is the time required for the source flux to change by
50\%. Recalling that we do not see variability on time intervals less
than a day, a $\sim$25\% change over 20 days, and a factor $\sim$2
over 2.5 months it is probably reasonable to assume that the
characteristic time scale is of order 1 month, give or take a factor
$\sim$2.
 
The 1-month characteristic time scale for variability (neglecting any
time dilation corrections for relativistic bulk motion) and the finite
light travel time of the escaping radiation impose an upper limit on
the size of the emitting region, $l \leq c\tau = 7.8\times10^{16}$ cm.
Compared with the independently estimated gravitational radius of
$R_{\rm g} = GM_{\rm BH}/c^{2} = 3.7\times10^{14}$ cm for the M87
nuclear black hole we obtain $l \lesssim 200 R_{\rm g}$. Thus the
processes responsible for the observed variability must occur in the
immediate vicinity of the central black hole. 

The total energy output in the variable component is another important
characteristic of the source. Our time sampling is too sparse to allow
an accurate estimate of the quiescent state, but the observed changes
of $\sim$2 indicate that at least half of the nuclear flux is
variable.  Integrating the observed spectrum at its highest state
between $100\mu$ and 10 Kev, gives a nuclear continuum luminosity of
order $\sim 3.0\times10^{42}$ ergs s$^{-1}$. This is only a small
fraction of the mechanical luminosity derived for the jet (Bicknell \&
Begelman 1996) which is $\sim 10^{44}$ ergs s$^{-1}$.  Even so, the
total power released by the BH in M87 is only a very small fraction of
the Eddington limit, $L_{\rm tot} \sim 10^{-4}L_{\rm Edd}$, indicating
that the central engine operates in a highly sub-Eddington regime. The
low optical luminosity and the sub-Eddington accretion supports the
advection dominated accretion flow (ADAF) model advocated for this
object by Reynolds et al.\ (1996). 

Three possible explanations for the observed variability are i) tidal
disruption of a star falling into the black hole (Hills 1975), ii)
instabilities in the relativistic accretion disk (Sunyaev 1973;
Shakura \& Sunyaev 1973), and iii) jet related processes (Kollgaard
1994, Camenzind \& Krockenberger 1992).

Although dwarf stars will cross the event horizon before being tidally
disrupted, giant stars with densities $\sim7\times10^{-5}$~g~cm$^{-3}$
can be disrupted at a distance of $\sim 7 R_{\rm g}$ from the M87
black hole.  The time scale for sudden disruption is a few days, and
the orbital period of the disrupted material is about two months in
either a Schwarzschild or Kerr metric. The expected frequency of tidal
disruptions is $\sim 10^{-3}$~yr$^{-1}$.  Given this low frequency,
continuing variability will argue against tidal disruption of stars.
 
If the variability is caused by instabilities in a ``classical''
accretion disk (e.g., hot spots), the variations in the observed flux
are caused by relativistic boosting by the radial component of the
orbital motion. In this case the characteristic time scale is
approximately given by the orbital period, and the variability will be
quasi-periodic depending on the spot life time. The shortest
rotational period around a $2.5 \times 10^9 M_{\odot}$ black hole is
$\sim$17.3 days in the Schwarzschild metric and $\sim$2.3 days in the
Kerr metric.  Hence, a detection of quasi-periodic variability on time
scales shorter than $\sim$17 days would argue against a nonrotating
black hole, and time scales shorter than $\sim$2 days would basically
exclude the accretion disk as the source of the observed
variability. Whatever the geometry, however, a very bright spot
relative to the rest of the disk is needed to reproduce the observed
amplitude.

In ADAF models the advective flow will suppress variability due to
relativistic effects close to the event horizon. However, variations
in the accretion rate may power variations in the relativistically
boosted jet.

Superluminal motion is often observed at the base of relativistic
radio jets. M87 is no exception (Biretta, Zhou \& Owen 1995).  The
sudden brightening and apparent motion is attributed to injection of
relativistic plasma into the jet. If the injected material carries
angular momentum from the accretion disk, it may follow a helical
trajectory and produce a quasi-periodic ``lighthouse effect'' because
of Doppler beaming and boosting (Camenzind \& Krockenberger 1992).
This model successfully explains the large simultaneous flux
variations observed in some blazars over many decades in frequency
from GHz to X-rays (Schramm et al.\ 1993, Wagner et al.\ 1995).
 
In order for the lighthouse effect to work in  M87, at least some
plasma in the jet must be moving very close to the line of sight.
Given that the inclination of the M87 jet is $\sim 30$\deg, and that
the opening angle of the inner jet is $\sim 6$\deg\ (Biretta 1994) such
a relativistic boosting can only occur if it is generated very close to
the base of the jet in the region where the jet is becoming
collimated.

We note that from the ground the usual seeing conditions make it very
difficult to measure reliably the variability of the relatively faint
nucleus against the stellar background.  The current observations can
not distinguish between variability originating in the accretion
disk and/or variability in the base of the jet enclosed by the FOS
aperture ($r \le 0.13''$; $r \le 9$ pc).  However, further
monitoring with appropriate sampling and polarization measurements may
identify the source of variability.  Nonetheless, these observations
have established that M87 displays yet another BL Lac-like property.

In summary, M87 has now been shown to possess the following properties
in common with the BL Lacs: 

\begin{list}{}{\leftmargin 0pt \itemindent 3em
\listparindent \itemindent \itemsep 1pt \parsep 1pt} 
\item[$-$] it is an elliptical galaxy
\item[$-$] it is a FR~I radio source
\item[$-$] it is a LINER 
\item[$-$] it has a relativistically boosted jet
\item[$-$] it displays superluminal motions in the jet
\item[$-$] it has a strongly variable nucleus, both at optical and
		X-ray frequencies
\item[$-$] it has an underlying featureless power law continuum spectrum 
\item[$-$] its spectral index is typical of BL Lacs
\end{list}

Unlike typical BL Lacs, M87 has a faint nucleus relative to the host
galaxy and would not be detected as a BL Lac if it were much further
away. Our results are consistent with the idea that the faintness of
the nucleus is primarily due to the relatively large angle between the
jet axis and the line of sight.
 
\acknowledgments 
We thank R.\ Sunyaev for insights into the variability timescales of
accretion disks and the referee for his comments which have helped us
to improve the presentation.  This work is based on observations with
the NASA/ESA Hubble Space Telescope, obtained at the Space Telescope
Science Institute, which is operated by the Association of
Universities for Research in Astronomy, Inc., under NASA contract
NAS5-26555. Support for the research was provided by NASA through
Grant NAG-1640 to the $HST$ FOS team.

\clearpage

 
\centerline{\bf FIGURE CAPTIONS}
 
\figcaption[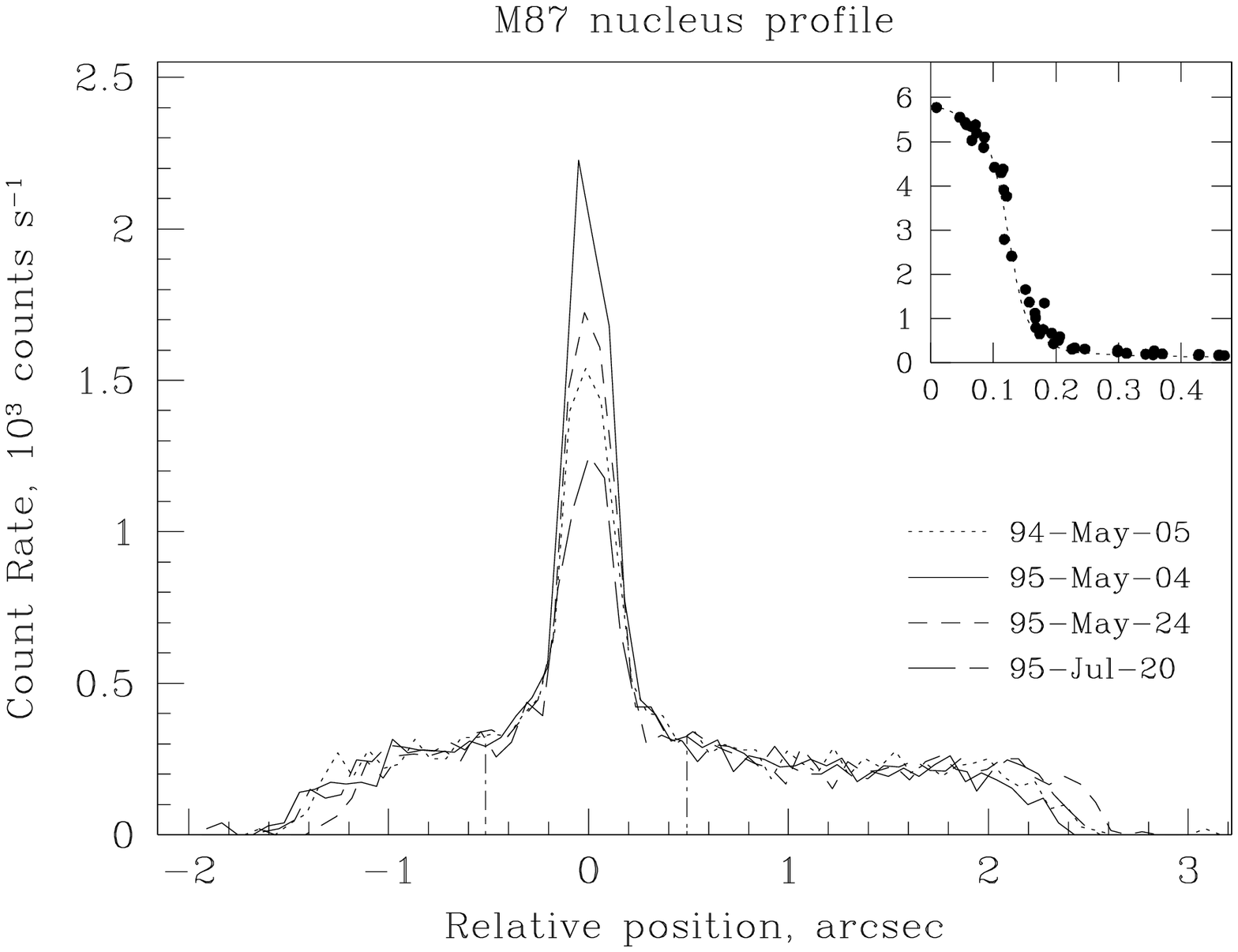] 
	{Brightness profiles of the nuclear region of M87 obtained
	from the binary search target acquisition data. Shown are four
	epochs of observation encompassing the observed range of
	variations.  The other two epochs (not shown for clarity) show
	comparable variation.  Each line represents a one dimensional
	image profile of the 1\farcs29$\times$3\farcs7 aperture
	projected onto the galactic center. The images are shifted, so
	that the centers of the nuclear profiles are aligned.  The two
	vertical dashed lines indicate the 1\farcs01 region used for
	estimating the nuclear flux.  The insert represents the radial
	profile of the peakup series for the 4 May 1995
	observation. The dashed line is a scaled profile of a star
	obtained in a peakup series through the same 0\farcs26
	circular aperture.}

\figcaption[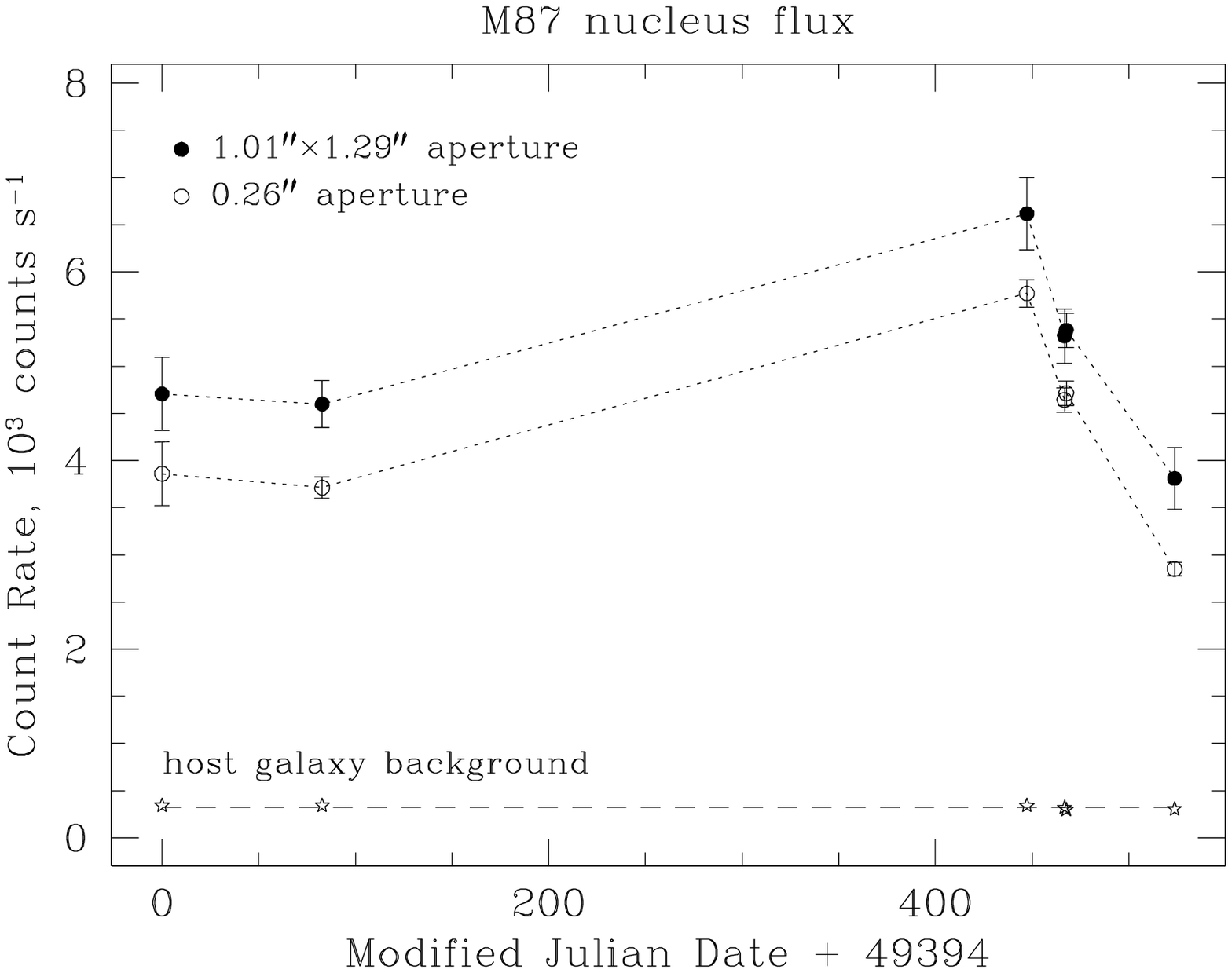]
	{Time variation of the nuclear flux measured from the binary
	search data (filled circles) and from the peakup series (open
	circles). The 1$\sigma$ error bars are about 5\% and 2.5\% for
	the BS and PU data, respectively. The error bars for the
	galactic component (stars) are smaller than the symbols and
	are not plotted.  The constant flux ratio between the
	0\farcs26 and 1\farcs01 $\times$ 1\farcs29 apertures is close
	to what is expected for the relative transmission of a point
	source (Keyes et al.\ 1995).  }

\figcaption[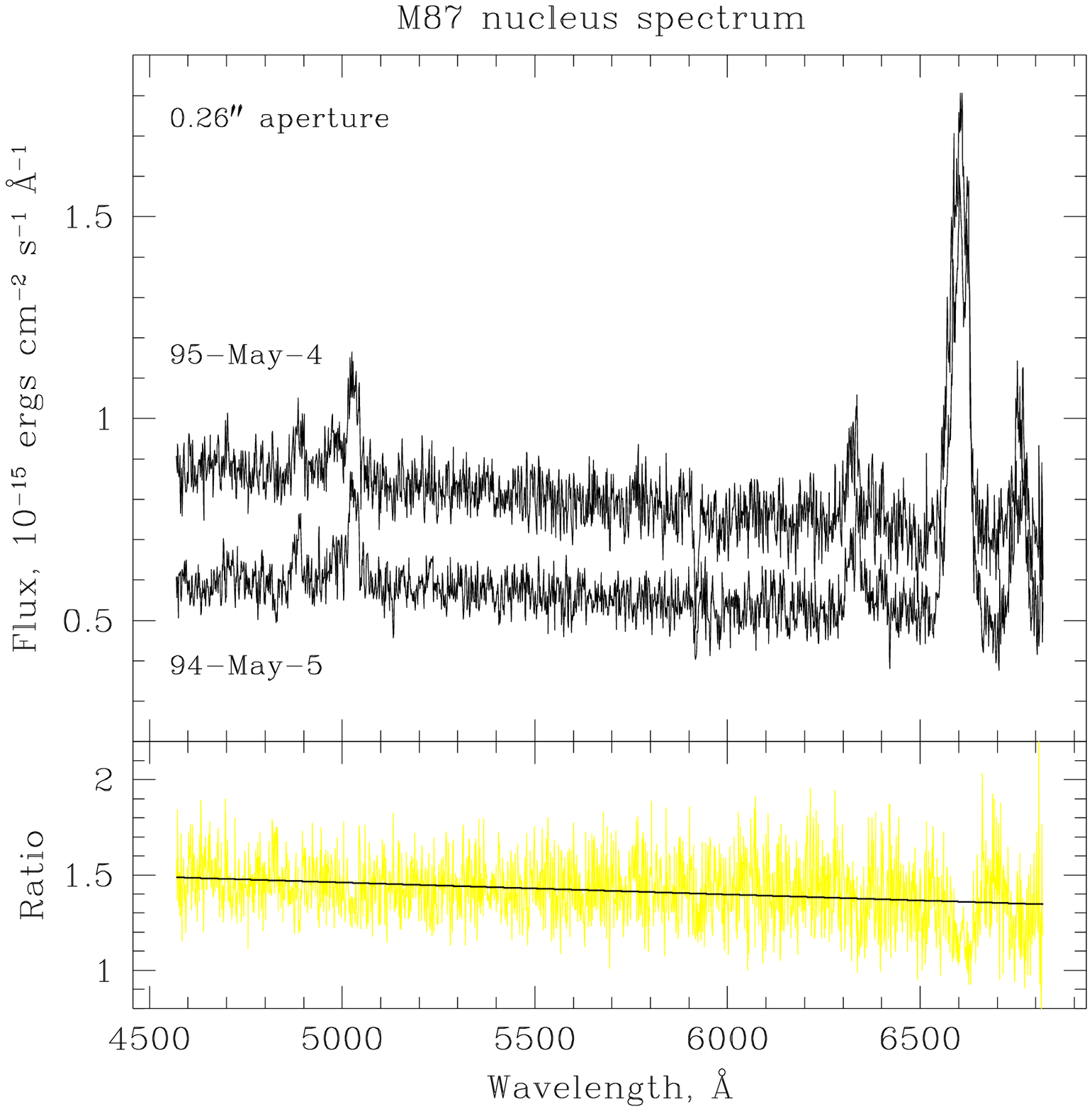]
	{The two spectra of M87's nucleus obtained through the
	0\farcs26 circular aperture.  Each spectrum has been smoothed
	with a 3-pixel boxcar to better represent the true
	resolution. The ratio of the two spectra is shown in light
	gray in the lower panel.  The continuous dark line is a simple
	linear regression fitted to the ratio.  Note the dips at the
	positions of the emission lines.}



\begin{figure}
\figurenum{1}
\plotone{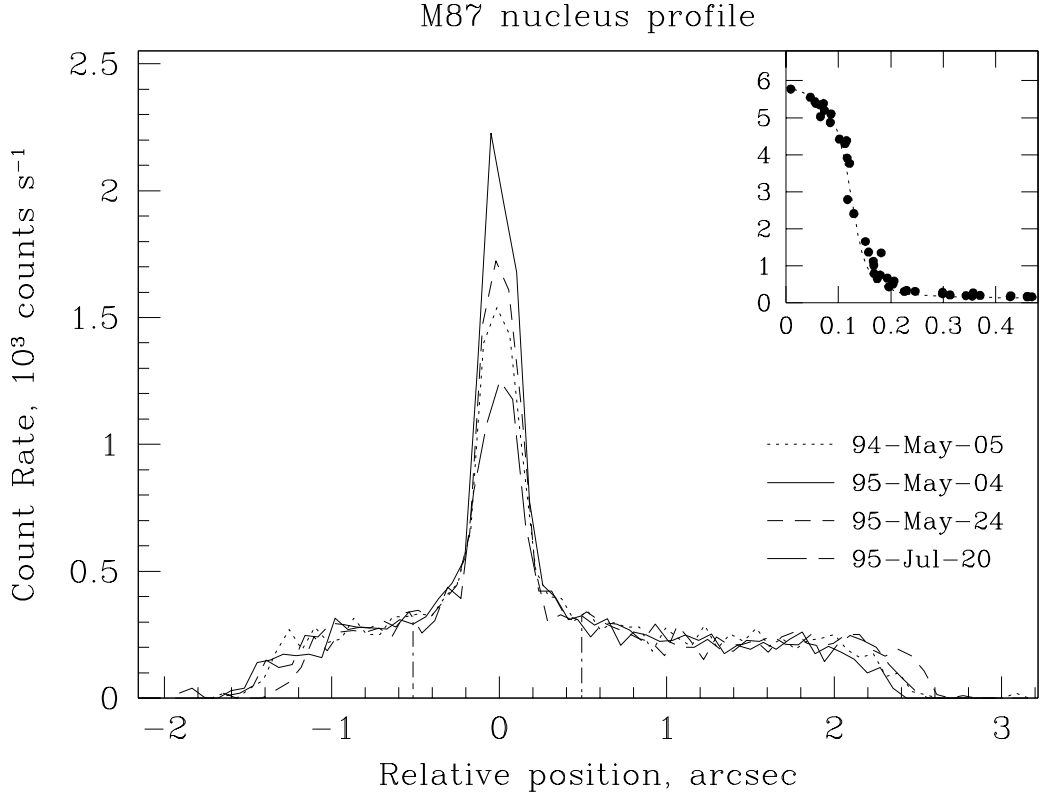}
\caption{Brightness profiles of the nuclear region of M87 obtained
from the binary search target acquisition data. Shown are four epochs
of observation encompassing the observed range of variations.  The
other two epochs (not shown for clarity) show comparable variation.
Each line represents a one dimensional image profile of the
1\farcs29$\times$3\farcs7 aperture projected onto the galactic
center. The images are shifted, so that the centers of the nuclear
profiles are aligned.  The two vertical dashed lines indicate the
1\farcs01 region used for estimating the nuclear flux.  The insert
represents the radial profile of the peakup series for the 4 May 1995
observation. The dashed line is a scaled profile of a star obtained in
a peakup series through the same 0\farcs26 circular aperture.}
\end{figure}
 
\begin{figure}
\figurenum{2}
\plotone{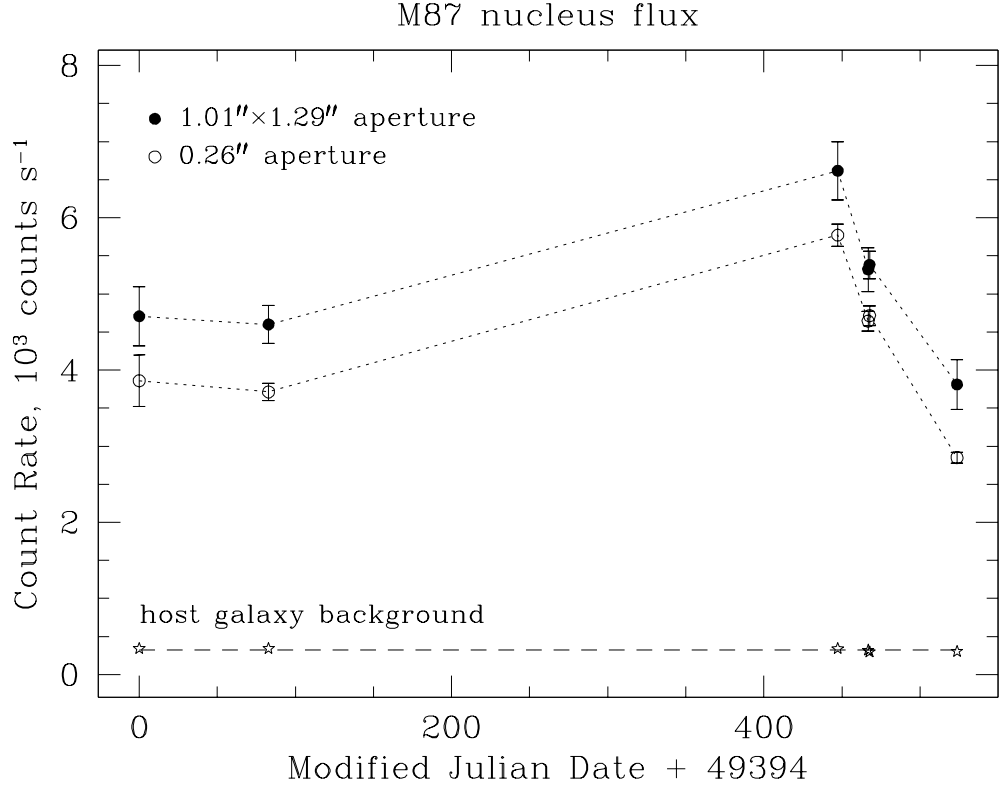}
\caption{Time variation of the nuclear flux measured from the binary
search data (filled circles) and from the peakup series (open
circles). The 1$\sigma$ error bars are about 5\% and 2.5\% for the BS
and PU data, respectively. The error bars for the galactic component
(stars) are smaller than the symbols and are not plotted.  The
constant flux ratio between the 0\farcs26 and 1\farcs01 $\times$
1\farcs29 apertures is close to what is expected for the relative
transmission of a point source (Keyes et al.\ 1995).  }
\end{figure}
 
\begin{figure}
\figurenum{3}
\plotone{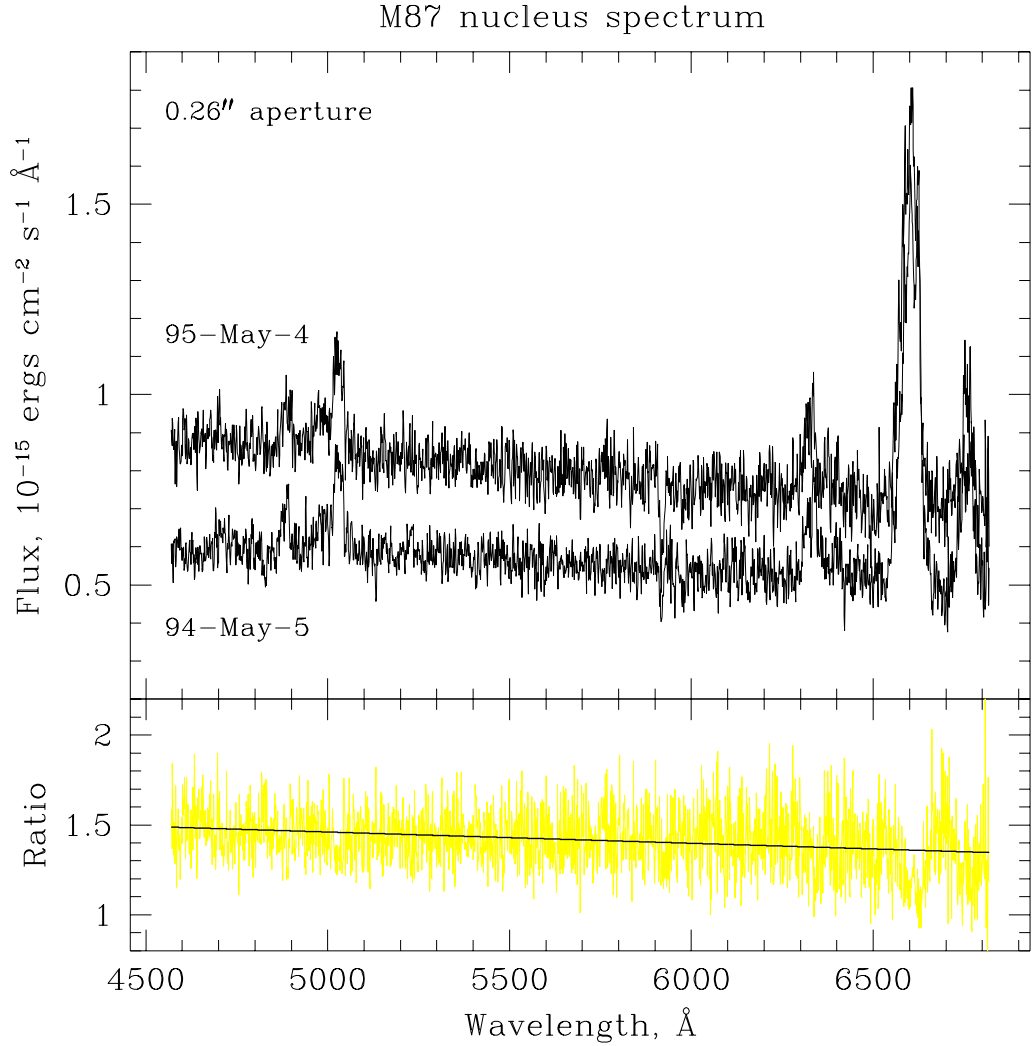}
\caption{The two spectra of M87's nucleus obtained through the
0\farcs26 circular aperture.  Each spectrum has been smoothed with a
3-pixel boxcar to better represent the true resolution. The ratio of
the two spectra is shown in light gray in the lower panel.  The
continuous dark line is a simple linear regression fitted to the
ratio.  Note the dips at the positions of the emission lines.}
\end{figure}

\end{document}